\documentclass[preprint]{elsarticle}
\usepackage{lineno,hyperref,pdflscape}
\modulolinenumbers[5]

\usepackage{multicol}
\usepackage{color}
\usepackage{xspace}
\usepackage{enumerate}
\usepackage{amsmath} 
\usepackage{makecell}
\usepackage{ulem} 
\usepackage{placeins}
\usepackage{threeparttable}

\usepackage{tabularx}
\usepackage{mathtools} 
\usepackage{amsfonts}
\newcolumntype{M}{>{\centering\arraybackslash}m{1.85cm}}
\usepackage[export]{adjustbox}
\usepackage{float}
\usepackage{braket}
\usepackage{lipsum}
\usepackage{longtable}
\usepackage{lscape}
\graphicspath{{figure/}}

\usepackage{booktabs}
\usepackage{blindtext}
\makeatletter
\newcommand{\colorcaption}[2][]{%
	\begingroup%
	\renewcommand{\@caption@fignum@sep}{ (Color online). }%
	\caption[#1]{#2}%
	\endgroup%
}

\makeatother

\usepackage{amssymb}
\usepackage{graphicx}
\usepackage{rotating}
\usepackage{tikz}
\begin{document}
	
	\begin{frontmatter}
		\title{  Large-scale shell-model study of two-neutrino double beta decay  of $^{82}$Se, $^{94}$Zr, $^{108}$Cd, $^{124}$Sn, $^{128}$Te, $^{130}$Te, $^{136}$Xe, and $^{150}$Nd}

		\author{Deepak Patel$^{1}$, Praveen C. Srivastava$^{1}$,\footnote{Corresponding author: praveen.srivastava@ph.iitr.ac.in}, V.K.B. Kota$^{2}$ and R. Sahu$^{3}$}
		\address{$^{1}$Department of Physics, Indian Institute of Technology Roorkee, Roorkee 247667, India}
		\address{$^{2}$Physical Research Laboratory, Ahmedabad 380 009, India}
		\address{$^{3}$National Institute of Science and Technology, Palur Hills, Berhampur-761008, Odisha, India}
		
		\date{\hfill \today}
		\begin{abstract}
			
			Large-scale shell-model calculations have been performed for the study of two-neutrino double-beta ($2\nu\beta\beta$) decay in $^{82}$Se, $^{94}$Zr, $^{108}$Cd, $^{124}$Sn, $^{128}$Te, $^{130}$Te, $^{136}$Xe, and $^{150}$Nd. We have employed JUN45 interaction to calculate the nuclear matrix element (NME) for $2\nu\beta\beta$ decay in $^{82}$Se. In the case of $^{94}$Zr, the glekpn effective interaction is used.  For $^{108}$Cd, we have used a realistic effective interaction derived through the G-matrix approach. In the case of $^{124}$Sn, $^{128,130}$Te and $^{136}$Xe, the sn100pn effective interaction is employed. For $^{150}$Nd, we have used KHHE effective interaction based on holes in a $^{208}$Pb core. We have extracted the half-lives of these nuclei for the $2\nu\beta\beta$ decay with the help of calculated NME. Our results are consistent with the available experimental half-lives. The variation of cumulative $2\nu\beta\beta$ NME with respect to the excitation energy of the intermediate $1^+$ states is also shown, and in all cases, it is ensured that their values are almost saturated.
			In the present work we have calculated more intermediate $1^+$ states as much as possible in comparison to results available in the literature.
			
		\end{abstract}
		
		\begin{keyword}
			Shell-model, Effective Interactions. 
		\end{keyword}
	\end{frontmatter}
	
	
	\section{Introduction}
	
	Double-beta decay (DBD) is the rarest radioactive decay process which is the subject of extensive study in the field of nuclear physics; see for example \cite{Ejiri, Vogel, Shimizu, Suhonen1, Toivanen, Caurier, Caurier1, Pritychenko1, Sahu1, Kostensalo, Sahu, Pacearescu, Barea, Kota1, Kostensalo1, Horoi, Yousef, Bobyk, Kotila, Yoshida, Nomura}. This decay process can be classified by two major decay modes: two-neutrino ($2\nu$) and neutrinoless ($0\nu$) double beta decay. The two-neutrino DBD ($2\nu\beta\beta$) involves the emission of two neutrinos and is allowed by conservation laws as a second-order weak process. The concept of $2\nu\beta\beta$ decay was first introduced by Mayer \cite{Mayer}. At present, this process has been observed in many nuclei in different mass regions of the nuclear chart \cite{Tretyak, Pritychenko, Barabash}. The observation of $2\nu\beta\beta$ decay provides experimental evidence of the standard model of particle physics and also provides good tests of nuclear models. This process also confirms the validity of the weak nuclear force and the existence of neutrinos as weakly interacting particles. On the other hand, $0\nu\beta\beta$ decay does not involve the emission of any neutrinos and violates the law of lepton number conservation. This exotic nuclear decay mode
	is a subject of many current experiments, and it is yet to be observed \cite{RMP}. It is important to add that the half-life measurements of $2\nu\beta\beta$ and $0\nu\beta\beta$ decay provide crucial information about the decay rates and lifetimes of isotopes involved in stellar nucleosynthesis, which impact the production of heavy elements in stars \cite{Zuber}. In this paper, we consider only $2\nu \beta\beta$ decay
	mode.
	
	Many different groups performed experiments measuring half-lives of double-beta decaying nuclei. Balysh $et$ $al.$ \cite{Balysh} observed the half-life of $2\nu\beta\beta$ decay candidate $^{48}$Ca, with $t_{1/2}^{2\nu}=(4.3^{+2.4}_{-1.1}[stat]\pm1.4[syst])\times 10^{19}$ yr. The half-life of $^{76}$Ge has been extracted as $t_{1/2}^{2\nu}=(1.84)^{+0.14}_{-0.10}\times 10^{21}$ yr in the Germanium Detector Array (GERDA) experiment \cite{GERDA}. In another experiment, the $2\nu\beta\beta$ decay half-life of $^{96}$Zr is obtained, $t_{1/2}^{2\nu}=[2.35\pm0.14(stat)\pm0.16(syst)]\times10^{19}$ yr using the NEMO-3 detector \cite{NEMO-3}. Similarly, using a liquid argon ionization chamber $2\nu\beta\beta$ decay half-life for $^{100}$Mo was determined in \cite{Barb-mo100} and combining with other experimental results, the average value obtained is $t_{1/2}^{2\nu}=(8.0 \pm 0.7) \times 10^{18}$ yr. Also, more recently Augier $et$ $al.$ \cite{Augier} investigated $2\nu\beta\beta$ and $0\nu\beta\beta$ decays of $^{100}$Mo to the first $0^+$ and $2^+$ excited states of $^{100}$Ru with the full CUPID-Mo exposure. Bernatowicz $et$ $al.$ \cite{Bernatowicz} confirmed the double-beta decay of $^{128}$Te and determined the ratio of half-lives for $\beta\beta$ decay of $^{130}$Te and $^{128}$Te as $T^{130}_{1/2}/T^{128}_{1/2}=(3.52\pm0.11)\times 10^{-4}$ using ion-counting mass spectrometry of Xe in ancient Te ores. Their calculated half-lives of $^{128}$Te, and $^{130}$Te are $(7.7\pm0.4\times)\times 10^{24}$ and $(2.7\pm0.1)\times 10^{21}$ yr, respectively. Recently, Alduino $et$ $al.$ \cite{Alduino} studied the $2\nu\beta\beta$ decay in $^{130}$Te using the CUORE-0 detector and determined $t_{1/2}^{2\nu}=[8.2\pm0.2(stat)\pm0.6(syst)]\times10^{20}$ yr. In the KamLAND-Zen $\beta\beta-$decay experiment \cite{Gando}, the $2\nu\beta\beta$ decay half-life of $^{136}$Xe is measured, $t_{1/2}^{2\nu}=2.38\pm0.02(stat)\pm0.14(syst)\times 10^{21}$ yr.
	
	Following the experimental works, several theoretical calculations have been performed for studying $2\nu$ DBD using many different nuclear models \cite{Kostensalo, Sahu, Pacearescu, Barea, Kota1, Kostensalo1, Horoi, Yousef, Bobyk, Nomura}. These nuclear models are used for obtaining the nuclear matrix elements (NME) needed for the analysis (see Eqs. (1) and (2) ahead).
	Firstly, the nuclear shell model is the most important theoretical framework to calculate the NMEs for DBD \cite{Kostensalo1, Horoi, Kota1}. Similarly, the many-body theoretical approach quasiparticle random-phase approximation (QRPA) is widely used for obtaining detailed predictions for $2\nu$ DBD half-lives \cite{Yousef, Bobyk}. The interacting boson model (IBM) is also used for calculating the NMEs for $2\nu\beta\beta$ decay \cite{Yoshida, Nomura}. Using these and many other nuclear models, one can predict the half-lives of double beta decaying nuclei where experimental data are unavailable.
	
	In the present work, we have performed systematic large-scale shell-model calculations for studying the $2\nu\beta\beta$-decay of $^{82}$Se, $^{94}$Zr, $^{108}$Cd, $^{124}$Sn, $^{128}$Te, $^{130}$Te, $^{136}$Xe, and $^{150}$Nd. We have extracted the half-lives of these nuclei using our calculated NMEs and compared them with the recent experimental data. Also, the variation of cumulative $2\nu\beta\beta$ NME with respect to the excitation energy of the intermediate $1^+$ states (see Eq. (2) ahead) is also studied.
	It is worthwhile to include the contribution of the maximum $1^+$ states of the intermediate nucleus up to the saturation level. This has been done in the present paper. These represent a comprehensive set of shell model results for the nuclei considered. 
	
	This paper is organized into the following sections. In Section \ref{section2}, the formalism for calculating the NMEs and half-lives is described briefly. The details about the interactions used in our calculations are also discussed. In Section \ref{section3}, we present our calculated results for the NMEs and half-lives of $2\nu\beta\beta$-decaying nuclei. The variation in the cumulative $2\nu\beta\beta$ NMEs with respect to excitation energy of the intermediate $1^+$ states is also discussed. Finally, we summarize our results and conclude the paper in section \ref{section4}.
	
	\section{Theoretical Framework} \label{section2}
	
	The half-life for the $2\nu\beta\beta$ decay can be expressed as follows
	
	\begin{equation}
		t_{1/2}^{2\nu}=\frac{1}{G^{2\nu}g_A^4|M_{2\nu}|^2},
		\label{eq1}
	\end{equation}
	
	where $G^{2\nu}$ is the phase-space factor \cite{Neacsu, Stoica, Kotila} and $g_A$ corresponds to axial-vector coupling strength. The nuclear matrix element (NME) $M_{2\nu}$ for $2\nu\beta\beta$ decay is given by
	
	\begin{equation}
		M_{2\nu}=\sum_{k}\frac{\langle 0_{g.s.}^{(f)}||\sigma\tau^{\pm}||1_k^+\rangle \langle 1_k^+||\sigma\tau^{\pm}||0_{g.s.}^{(i)}\rangle}{[\frac{1}{2}Q_{\beta\beta}+E(1^+_k)-M_i]/m_e+1},
	\end{equation}
	
	where $m_e$ is the rest mass of the electron; $E(1^+_k)-M_i$ denotes the energy difference between the $k^{th}$ intermediate $1^+$ state and the ground state of the initial nucleus; $0_{g.s.}^i$($0_{g.s.}^f$) is the ground state of initial (final) nuclei; $\sigma$ is the pauli matrix; $\tau^-(\tau^+)$ is the isospin lowering (raising) operator. $Q_{\beta\beta}$ ($Q$-value) is the energy released in the decay. $\langle 0_{g.s.}^{(f)}||\sigma\tau^{\pm}||1_k^+\rangle$ (or $\langle 1_k^+||\sigma\tau^{\pm}||0_{g.s.}^{(i)}\rangle$) is the reduced matrix element and can be expressed as
	
	\begin{equation}
		\langle J_f||\sigma \tau^{\pm}||J_i \rangle = \sum_{j_f j_i} \sqrt{3(2j_f+1)}\delta_{l_il_f} U(l_is_ij_f1,j_is_f)D_{j_fj_i}.
	\end{equation}
	
	Here, the term $\sqrt{3(2j_f+1)}\delta_{l_il_f} U(l_is_ij_f1,j_is_f)$ denotes the reduced single-particle matrix elements \cite{Suhonen_book, Jia} for Gamow-Teller (GT) transition, the $U$-coefficient is Racah coefficient \cite{Edmonds} and $j_i$, $l_i$, and $s_i$ ($j_f$, $l_f$, and $s_f$) are the total angular momentum, orbital angular momentum, and spin of initial (final) nucleonic state, respectively. $\delta_{l_il_f}$ shows that for the allowed GT transition, the orbital angular momentum of the initial and final state nucleons must be equal. $D_{j_fj_i}$ represent the one-body transition densities and can be expressed as
	
	\begin{equation}
		D_{j_fj_i}=\frac{\langle f||a_{j_f}^{\dagger}a_{j_i}||i\rangle}{\sqrt{2\delta_j+1}},
	\end{equation}
	
	where $a_{j_f}^{\dagger}$ ($a_{j_i}$) is nucleon-creation (annihilation) operator, and $\delta_j$ represents the changing of the angular momentum.
	
	We have performed our calculations using shell-model code NuShellX \cite{Nushellx}. For the study of $2\nu\beta\beta$ decay in $^{82}$Se, the shell-model calculations have been conducted using JUN45 interaction \cite{Honma}, which consists of the $0f_{5/2}1p0g_{9/2}$ proton and neutron orbitals.  This interaction is developed by starting with a realistic interaction based on the Bonn-C potential and with an empirical modification in the single-particle energies of four orbitals and 133 two-body matrix elements. Here, 5000 intermediate $1^+$ states in $^{82}$Br are calculated up to the excitation energies of 24.874 MeV.
	
	In the case of $^{94}$Zr, we have used glekpn effective interaction \cite{Mach} having the $0f1p0g_{9/2}$ proton orbitals and $0g1d2s$ neutron orbitals. In our calculation, we have completely filled the proton orbital $0f_{7/2}$ and neutron orbital $0g_{9/2}$. We have calculated 150 intermediate $1^+$ states for $^{94}$Nb up to the excitation energies of 6.492 MeV.

	For the calculation of NME of $2\nu\beta\beta$-decay in $^{108}$Cd, we have used a realistic effective interaction obtained by the G-matrix approach, which was constructed by the effective microscopic interaction derived from a charge-symmetry breaking nucleon-nucleon potential \cite{Machleidt} with further modifications in the monopole part and was used in Ref.~\cite{Boelaert}. Here, we have included $1p_{1/2}0g_{9/2}$ proton orbitals and $0g_{7/2}1d2s$ neutron orbitals and excluded the $0h_{11/2}$ neutron orbital. In our calculation, 5000 intermediate $1^+$ states in $^{108}$Ag are calculated up to the excitation energies of 11.288 MeV.
	
	In the case of $^{124}$Sn, $^{128,130}$Te and $^{136}$Xe, the shell-model effective interaction sn100pn \cite{Brown1} is used for our calculations. This interaction consists of $0g_{7/2}1d2s0h_{11/2}$ orbitals for protons and neutrons between $50-82$ model space. This interaction is obtained by starting with a G-matrix derived from the CD-Bonn \cite{Machleidt1} nucleon-nucleon interaction. We have used full model space for $^{136}$Xe. In case of $^{124}$Sn, $^{128,130}$Te, we have excluded proton orbital $h_{11/2}$. Additionally, we have fully filled the lowest lying neutron orbital $g_{7/2}$ for $^{128}$Te. Here, we consider 3000 intermediate $1^+$ states for $^{128}$I and 5000 states for $^{130}$I, and $^{136}$Cs. The shell model dimension is relatively large for $^{124}$Sb, so we have calculated 100 intermediate $1^+$ states in this case.
	
	For $^{150}$Nd, we have performed shell model calculation using KHHE effective interaction \cite{Warburton1, Warburton2} in the model space with $Z$=50-82 and $N$=82-126. This interaction was constructed based on holes in the $^{208}$Pb core. The effective interaction of Kuo and Herling \cite{Kuo1, Kuo2} was derived from a free nucleon-nucleon potential of Hamada and Johnston \cite{Hamada} and further renormalized due to the finite extension of model space by the reaction matrix techniques developed by Kuo and Brown \cite{Kuo3}. This interaction consists of $0g_{7/2}$, $1d_{5/2}$, $1d_{3/2}$, $2s_{1/2}$, and $0h_{11/2}$ proton orbitals and $0h_{9/2}$, $1f_{7/2}$, $1f_{5/2}$, $2p_{3/2}$, $2p_{1/2}$, and $0i_{13/2}$ neutron orbitals. To make our calculation feasible, we have employed truncation where all the partitions belonging to $\pi({g_{7/2}^{4-8}d_{5/2}^{0-0}d_{3/2}^{0-0}s_{1/2}^{2-2}h_{11/2}^{0-12}})$ and $\nu({h_{9/2}^{4-10}f_{7/2}^{0-0}f_{5/2}^{0-0}p_{3/2}^{0-0}p_{1/2}^{0-0}i_{13/2}^{0-14}})$ configurations. 
	Using a similar kind of Hamiltonian (Kuo-Herling interaction) \cite{Kuo1,Kuo2}, Monte-Carlo shell model calculations  were also performed to study the $0\nu\beta\beta$ decay of $^{150}$Nd in Ref. \cite{Shimizu1}. In our calculation, 75 intermediate $1^+$ states in $^{150}$Pm are calculated up to the excitation energies of 2.773 MeV.

	\section{Results and Discussion} \label{section3}
	
	In this section, we have discussed the shell model predicted NMEs and half-lives of $^{82}$Se, $^{94}$Zr, $^{108}$Cd, $^{124}$Sn, $^{128}$Te, $^{130}$Te, $^{136}$Xe, and $^{150}$Nd isotopes for $2\nu\beta\beta$ decay process, the calculated results of nuclei of interest are reported in Table \ref{half-life}. The $Q_{\beta\beta}$ and $Q_{\beta}$ values are taken from the Ref. \cite{NNDC, Jia}.  In our calculations, the effective $g_A$ values are taken from Ref. \cite{Suhonen}, except for $^{108}$Cd, $^{124}$Sn, and $^{150}$Nd. In the case of these three isotopes, we have taken $g_A=1.00$. A careful discussion of NME ($M_{2\nu}$) is very important for the study of $2\nu\beta\beta$ decay of a nucleus. In Fig. \ref{fig1}, we have shown the variation of cumulative NME as a function of excitation energy of the intermediate $1^+$ states.
	
	\subsection{ \bf Variation of cumulative nuclear matrix elements (NMEs) and extracted half-life} 
	
	{\bf $^{82}$Se}: In case of $^{82}$Se, the shell-model calculated NME for $2\nu\beta\beta$ decay is $|M_{2\nu}|=0.1713$. For the precise calculation of the $2\nu\beta\beta$ NME, the exact energies of intermediate states play an important role. In our calculation, the excitation energies of the intermediate $1^+$ states in $^{82}$Br were shifted such that the lowest lying $1^+$ state is at the experimental energy of 0.075 MeV. The contribution of NME from the lowest $1^+$ state is 0.0557, which is 32.5\% of the total NME (0.1713). Up to the $10^{th}$ intermediate $1^+$ state, the cumulative NME is added constructively with a value of 0.1125. For the $11^{th}$ state, a small decrement is shown from 0.1125 to 0.1103. After that, the nature of cumulative NMEs is almost constructive up to 7.0 MeV. Beyond this, very small variations occur with a final value of 0.1713. Previously, the half-life ($t_{1/2}^{2\nu}$) of $^{82}$Se in $2\nu\beta\beta$ decay process has been calculated in different theoretical works \cite{Sahu, Dhiman, Caurier3} with different models. Using the shell-model calculated NME, the extracted $t_{1/2}^{2\nu}$ value (0.68$\times 10^{20}$ yr) is showing better agreement with the average value $0.87^{+0.02}_{-0.01}\times 10^{20}$ yr \cite{Barabash} than the previous works.

	\begin{landscape}
		\begin{table*}
			\centering
			\caption{Shell-model calculated $2\nu\beta\beta$ NMEs and the extracted half-lives.}
			\begin{tabular}{cccccccc}
				\hline\hline
				Isotope	& $|M_{2\nu}|$ & $G^{2\nu}$ (yr$^{-1}$)  & $g_{A}^{eff}$ \cite{Suhonen} & Calculated $t_{1/2}^{2\nu}$ (yr) &  \makecell{Experimental/Recommended \\ (Average) value of $t_{1/2}^{2\nu}$ (yr)}  \\
				\hline
				&  &   & & &  \\
				
				$^{82}$Se	& 0.1713  & 150.31$\times 10^{-20}$ \cite{Neacsu} & 0.76 & 0.68$\times 10^{20}$  & $0.87^{+0.02}_{-0.01}\times 10^{20}$  \cite{Barabash}  \\
				
				\hline
				&  &   & & &  \\
				
				$^{94}$Zr	& 0.0611  & 2.28$\times 10^{-21}$ \cite{Jia} & 0.60 & 9.08$\times 10^{23}$  & (0.31$\sim$66)$\times 10^{23}$ \cite{Tretyak}  \\
				
				\hline
				&  &   & & &  \\
				
				$^{108}$Cd	& 0.1729  & 2.803$\times 10^{-26}$  \cite{Raina} & 1.00 & 1.19$\times 10^{27}$  & (3.939$\sim$9.959)$\times 10^{27}$  \cite{Raina}  \\
				
				&  & &  & & $> 4.1\times10^{17}$ \cite{Danevich}  \\
				
				\hline
				&  &   & & &  \\
				
				$^{124}$Sn	& 0.0369  & 51.45$\times 10^{-20}$ \cite{Neacsu} & 1.00 & 1.43$\times 10^{21}$  & -  \\
				
				\hline
				&  &  & & &   \\
				$^{128}$Te  & 0.0255 & 8.401$\times 10^{-22}$ \cite{Semenov} & 0.72 & 6.82$\times 10^{24}$ & (2.25$\pm$0.09)$\times10^{24}$ \cite{Barabash} \\
				
				\hline
				&  &   & &  & \\
				$^{130}$Te	& 0.0516  & 142.73$\times 10^{-20}$ \cite{Neacsu} & 0.72 & 9.78$\times 10^{20}$  &  (7.91$\pm$0.21)$\times 10^{20}$ \cite{Barabash}  \\
				
				\hline
				&  &   & &  &  \\
				$^{136}$Xe	& 0.0748  & 133.73$\times 10^{-20}$ \cite{Neacsu} & 0.57 & 1.27$\times 10^{21}$  &  (2.18$\pm$0.05)$\times 10^{21}$ \cite{Barabash} \\
				
				\hline
				&  &   & &  &  \\
				$^{150}$Nd	& 0.0691  & 3467.53$\times 10^{-20}$ \cite{Neacsu} & 1.00 & 6.05$\times 10^{18}$  &  (9.34$\pm$0.65)$\times 10^{18}$ \cite{Barabash} \\
				
				\hline\hline	
				
			\end{tabular}
			\label{half-life}
		\end{table*}
	\end{landscape}

	{\bf $^{94}$Zr}: Our calculated NME for the $^{94}$Zr is $|M_{2\nu}|=0.0611$. There is no $1^+$ state confirmed experimentally for $^{94}$Nb. So, we have used shell-model excitation energies for the calculation of NME. The contribution in the total NME from the lowest $1^+$ state is 0.0364. The next three states are constructive, and the contribution of NME increases with a value of 0.0621. The next (5$^{th}$) state adds the cumulative NME destructively and brings to the value of 0.0596. Further, the next four states added the contribution in NME almost constructively to a value of 0.0611. After that, very small variations are shown in the NME. Using the final value of NME, phase-space factor ($G^{2\nu}$) \cite{Neacsu}, and the effective $g_A$ value \cite{Suhonen} in equation (\ref{eq1}), we have calculated the half-life of $^{94}$Zr for the $2\nu\beta\beta$ decay. In Ref. \cite{Tretyak}, the calculated half-life of $^{94}$Zr for $2\nu\beta\beta$ decay using the pnQRPA method lies in between (0.31$\sim$66)$\times 10^{23}$ yr. As reported in Table \ref{half-life}, our calculated $t_{1/2}^{2\nu}$ value is compatible with this previous calculation \cite{Tretyak}.

	{\bf $^{108}$Cd}: We have calculated the NME in $2\nu$ double-electron capture ($ECEC$) process for $^{108}$Cd, which equals 0.1729. This value is approximately 8\% larger than the previously calculated NME 0.160, in Ref. \cite{Jia}. The lowest intermediate $1^+$ state contributes above 50\% (0.0947) of the final value. The next two states are constructive with a value of 0.1704 (at 1.402 MeV). The NMEs cumulate constructively between 1.402 to 2.000 MeV (up to a maximum value of 0.1771). From the sixteenth intermediate state (at 2.072 MeV), the cumulative NME decreases slowly up to 6.378 MeV energy with a value of 0.1730. Hereafter, the cumulative NMEs are almost constant, with a final value of 0.1729. Experimentally, the lower limit for the half-life of $^{108}$Cd in $2\nu$ double-electron capture process is given by different groups \cite{Danevich, Kiel}. Using our calculated NME (0.1729), we obtained $t_{1/2}^{2\nu}=1.193\times 10^{27}$ yr. This value is close to the previous values calculated with the shell model \cite{Jia} and the projected Hartree-Fock-Bogoliubov (PHFB) model \cite{Raina}. In the half-life calculation, we have taken $g_A^{eff}=1$. Using a smaller $g_A^{eff}$ value, one can extract $t_{1/2}^{2\nu}$ within the range reported in Ref. \cite{Raina}.

	\begin{figure*}
		\includegraphics[width=59mm]{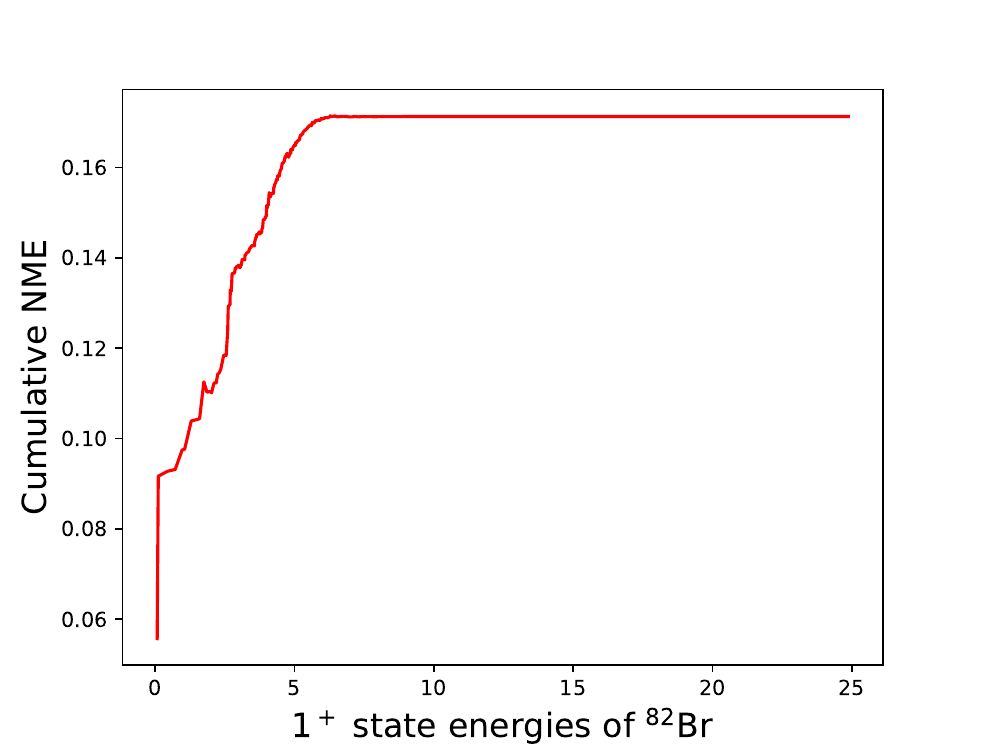}
		\includegraphics[width=59mm]{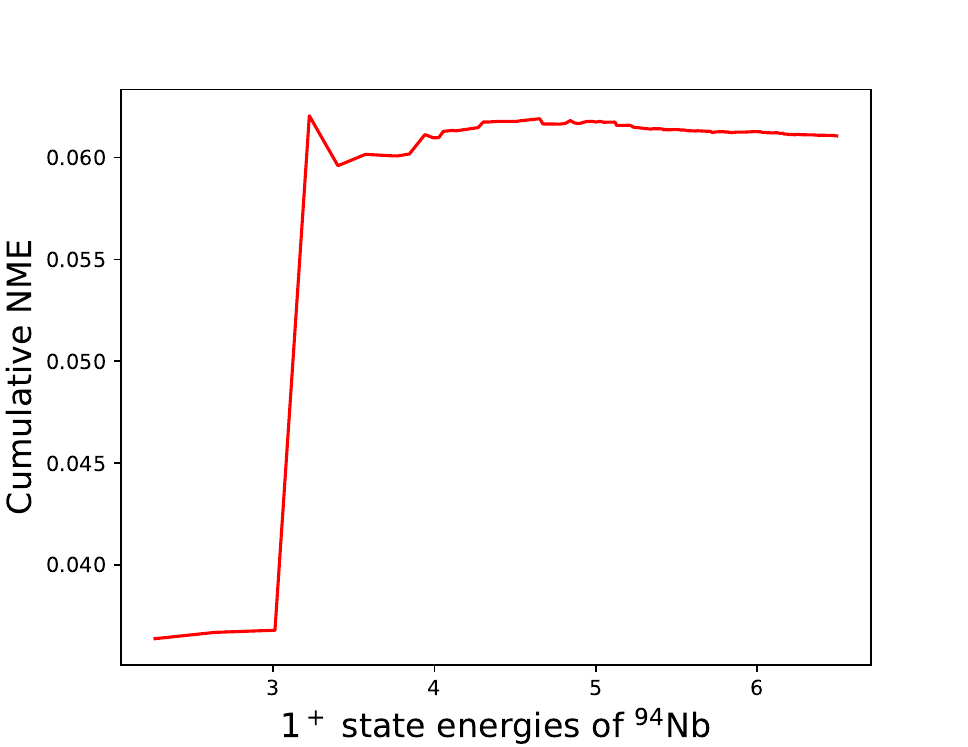}
		\includegraphics[width=59mm]{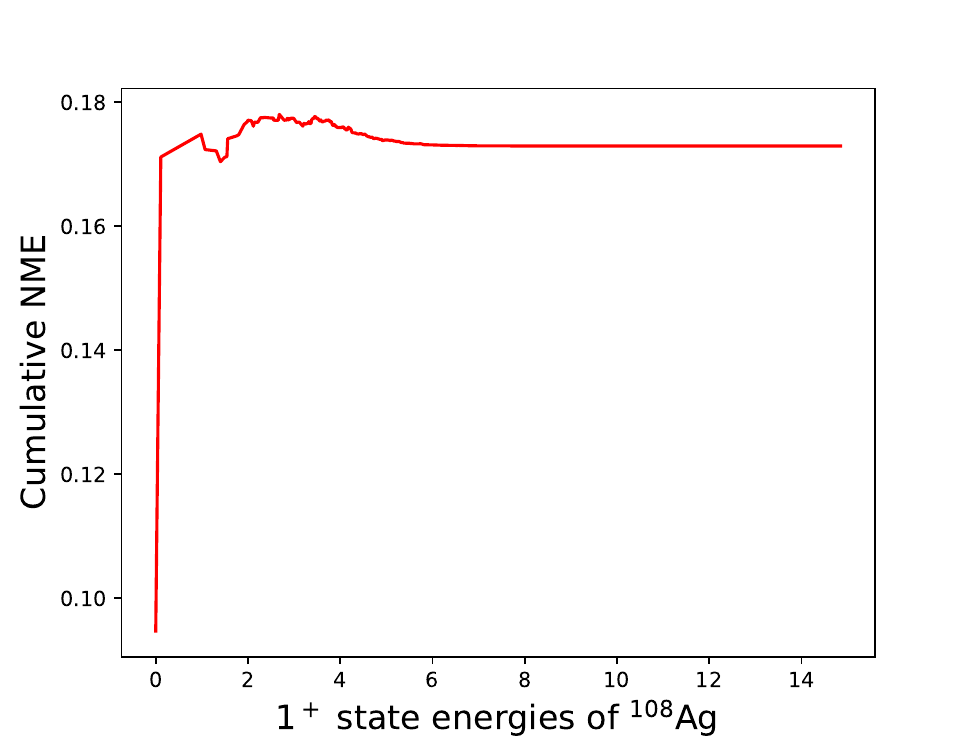}
		\includegraphics[width=59mm]{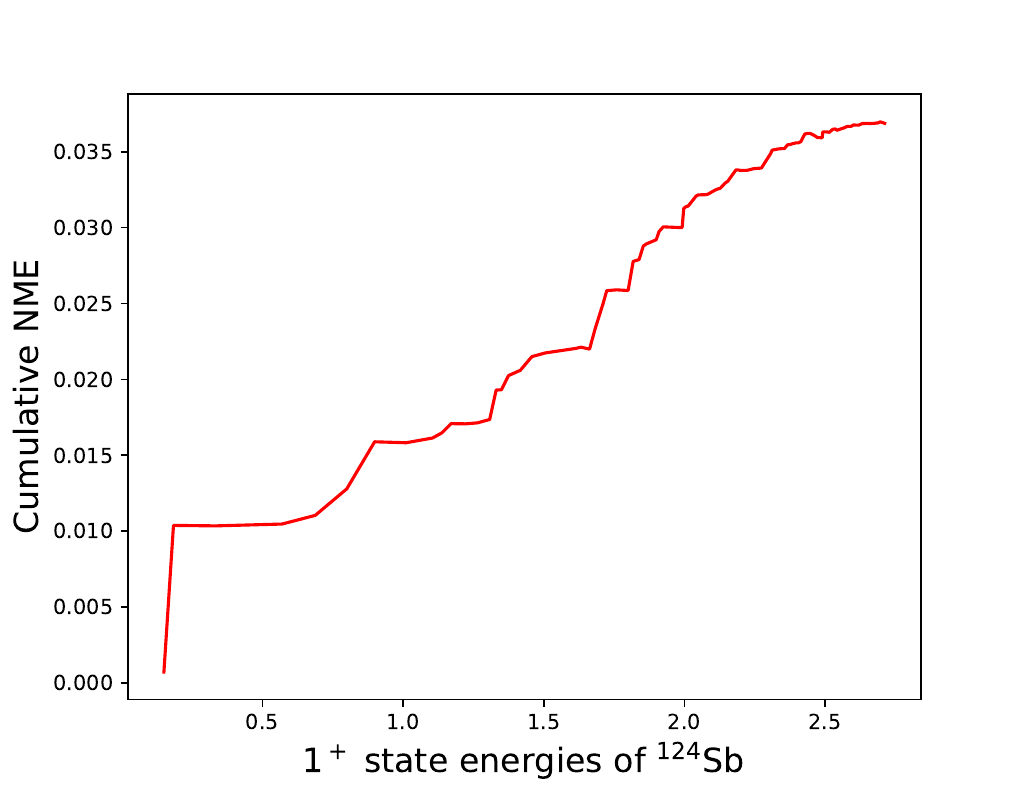}
		\includegraphics[width=59mm]{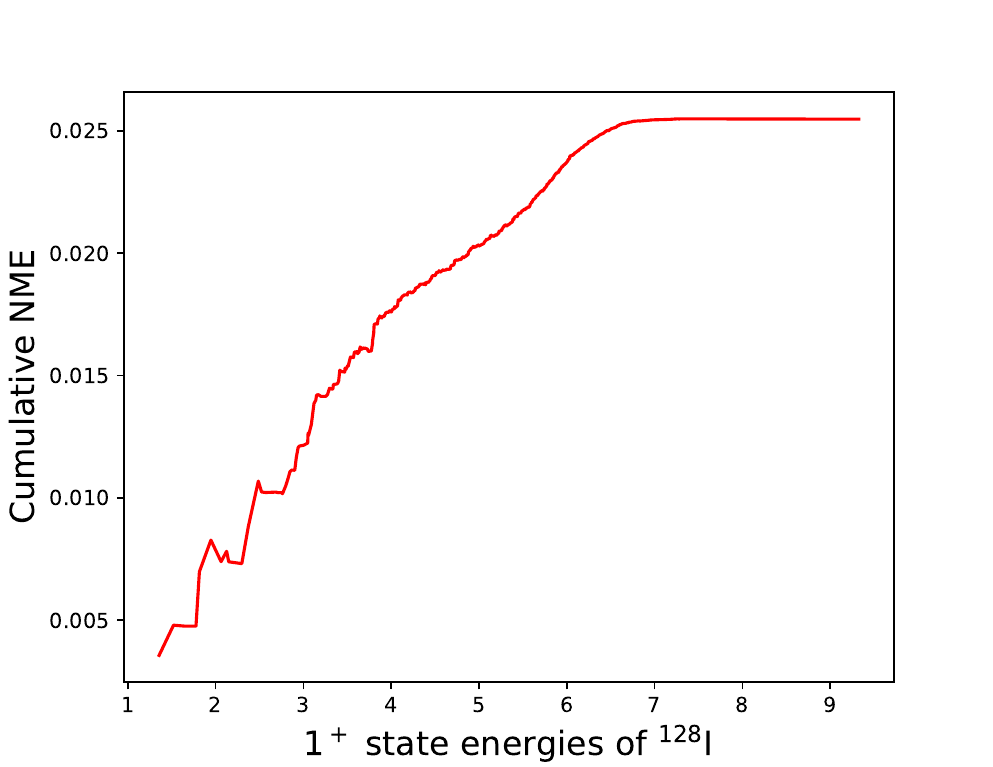}
		\includegraphics[width=59mm]{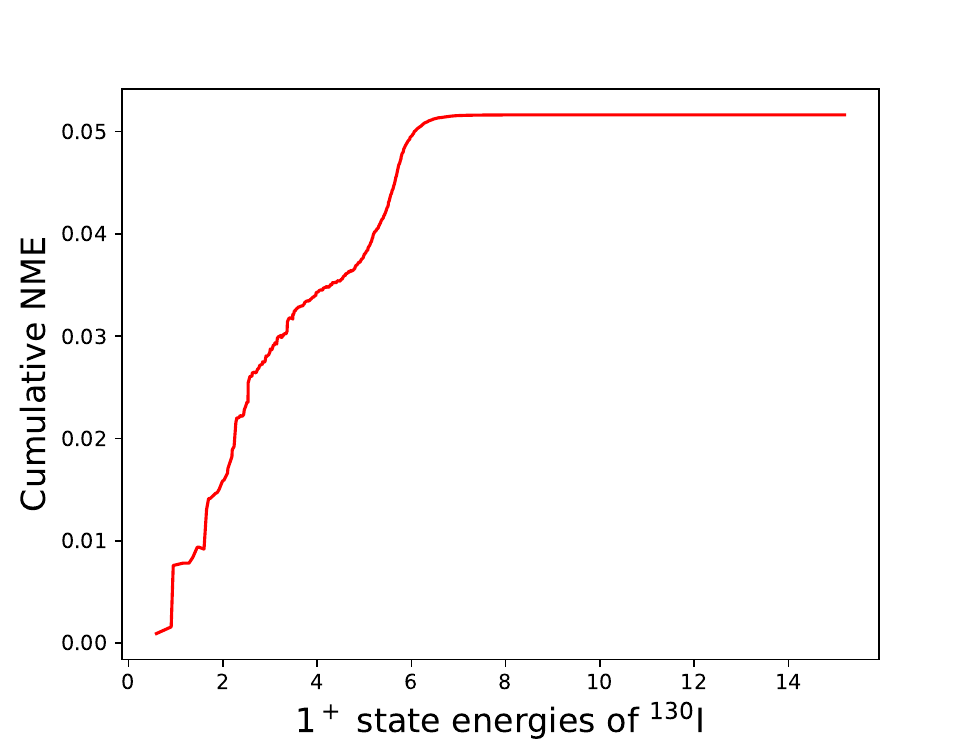}
		\includegraphics[width=59mm]{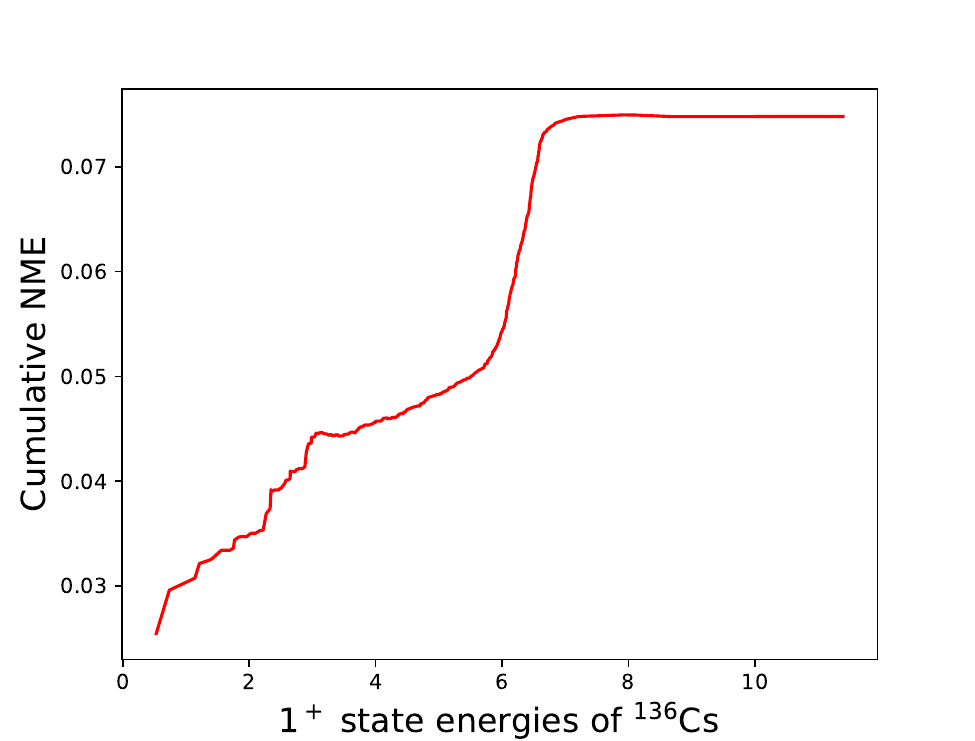}
		~~~\includegraphics[width=59mm]{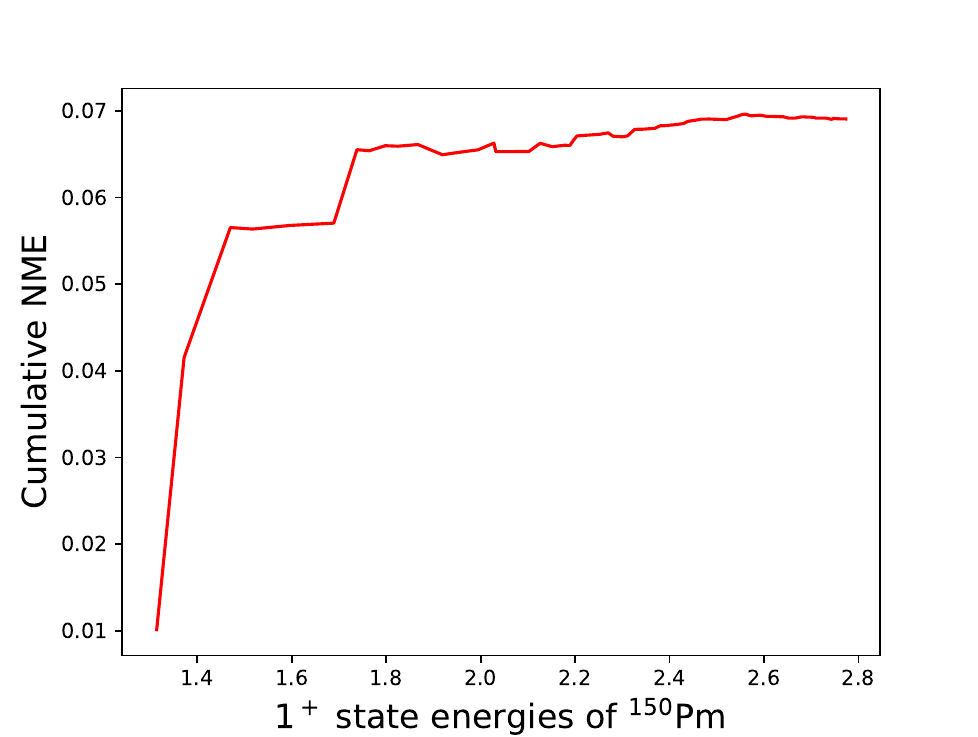}
		\caption{\label{fig1} Cumulative $2\nu\beta\beta$ NME ($M_{2\nu}$) for $^{82}$Se, $^{94}$Zr,  $^{108}$Cd, $^{124}$Sn, $^{128}$Te, $^{130}$Te, $^{136}$Xe, and $^{150}$Nd as a function of excitation energy of the intermediate state in $^{82}$Br, $^{94}$Nb, $^{108}$Ag, $^{124}$Sb, $^{128}$I, $^{130}$I, $^{136}$Cs, and $^{150}$Pm, respectively.}
	\end{figure*}

	{\bf $^{124}$Sn}:  In the case of $^{124}$Sn, the shell model predicted NME for $2\nu\beta\beta$ decay is 0.0367. In our calculation, we have adjusted the shell model calculated intermediate $1^+$ state energies in $^{124}$Sb such that the lowest lying $1^+$ state is shifted at the experimental energy of 0.150 MeV. The contribution of the lowest $1^+$ state in the final NME is very small (0.0007). Further, the next 99 intermediate $1^+$ states contribute almost constructively except for a few states within the energy range $\sim$2.712 MeV. Experimentally, the half-life of $^{124}$Sn for $2\nu\beta\beta$ decay is not observed yet. Using the final value of NME, $M_{2\nu}=0.0369$, we have extracted the half-life as $T_{1/2}^{2\nu}=1.43\times10^{21}$ yr. Our calculated half-life is close to the previous shell model obtained half-life $1.6\times10^{21}$ yr \cite{Horoi1}.
	
	{\bf $^{128}$Te}:  Our calculated $2\nu\beta\beta$ NME for $^{128}$Te is $|M_{2\nu}|=0.0255$, which is comparable to the value 0.0249 reported in the Ref. \cite{Barabash1}. The lowest $1^+$ state in $^{128}$I contributes a small amount of cumulative NME equals 0.0036. The next $1^+$ state cumulates the NME constructively, but the $3^{rd}$ and $4^{th}$ intermediate $1^+$ states show a destructive nature. Again, the next two consecutive states ($5^{th}$ and $6^{th}$) are constructive with a value of 0.0083. The $7^{th}$ intermediate $1^+$ state decreases the NME (0.0074), and for $8^{th}$ intermediate state, it increases to a value of 0.0078. Further, the next two consecutive states are again destructive. After that (at 2.373 MeV), the NME increases almost constructively except for some cases and starts to saturate approximately at 7.0 MeV (after 1900 states). Using the final value of NME, our calculated half-life of $^{128}$Te in $2\nu\beta\beta$ decay is 6.82$\times 10^{24}$ yr, which is close to the recommended value \cite{Barabash} (see in Table \ref{half-life}).
	
	{\bf $^{130}$Te}: In the case of $^{130}$Te, the shell-model calculated NME, $|M_{2\nu}|$ is 0.0516. The $1^+$ state is not confirmed also for $^{130}$I. From the lowest $1^+$ state, the contribution in the total NME is much less and equals 0.0009. With the next 2000 states (up to 6.78 MeV), the cumulative NME added constructively to a value of 0.0515. After the 2000 states, the variations in the NME are negligible, and the cumulative matrix element shows a constant nature after 4000 states (at 12 MeV) with a final value of 0.0516. Barabash \cite{Barabash} reported the experimental half-life of $^{130}$Te for $2\nu\beta\beta$ decay from different experimental works and suggested the average value of $t_{1/2}^{2\nu}$ is $(7.91\pm 0.21)\times 10^{20}$ yr. Our calculated $t_{1/2}^{2\nu}$ shows a good agreement with this value and the other experimental measurements.
	Present result confirming that we have good model space choice and  effective interaction giving reasonable agreement with data for $2\nu\beta\beta$ decay half-life, one can confidently use these inputs and calculate $0\nu\beta\beta$ decay NME so that we can obtain good bounds on neutrino mass. 
	
	{\bf $^{136}$Xe}: For $^{136}$Xe, the first excited $1^+$ state contributes an amount of 0.0255, which is 34.1\% of the total NME (0.0748). From the lowest $1^+$ state, cumulative NME almost increases consistently up to 80 states. But, after 3.616 MeV ($81^{th}$ state), three consecutive states are destructive with a small decrement. Beyond that, the cumulative NME slowly increases consistently. After the 550 states (at 6.0 MeV), the NME increases rapidly up to a value of 0.0745 (reaching 7.0 MeV). Beyond this, the cumulative NME varies slowly, with a final value of 0.0748. Using this NME value, we calculated the half-life of $^{136}$Xe for $2\nu\beta\beta$ decay equals $1.266\times 10^{21}$ yr. From Table \ref{half-life}, we can see that our calculated $t_{1/2}^{2\nu}$ value is close to the average value $(2.18\pm 0.05)\times 10^{21}$ in Ref. \cite{Barabash}.
	
	{\bf $^{150}$Nd}: In the case of $^{150}$Nd, the shell-model calculated NME for $2\nu\beta\beta$ decay is 0.0691. This value is close to the NME (from the recent compilation of Ref. \cite{Barabash1}) reported in Ref. \cite{Caurier1}. We have used the shell-model calculated $1^+$ states in $^{150}$Pm for the calculation of NME. The lowest $1^+$ state contributes an amount of 0.0101 in the total value of NME. The behavior of cumulative NMEs is dominantly constructive except for some states up to $75^{th}$ intermediate $1^+$ state. Using the final value of NME, we obtained half-life $t_{1/2}^{2\nu}=6.05\times 10^{18}$ yr, which is showing a good comparison with the recommended value $(9.34\pm0.65)\times 10^{18}$ yr \cite{Barabash}. Here, we are unable to calculate the NME beyond $75^{th}$ state because of the huge dimension. It could  be possible to get more improved results if we calculate more intermediate states. 
	
	In our calculation, we have considered the contribution of intermediate $1^+$ states up to very high excitation
	energies (see $^{82}$Br, $^{108}$Ag, $^{128}$I, $^{130}$I, and $^{136}$Cs in Fig. \ref{fig1}). Here, the question arises whether such a calculation to high excitation energies is realistic in view of the limited model space considered. In the shell model calculation of Caurier et al. \cite{Caurier1} for $\beta\beta$ decay, they have considered the contribution of $1^+$ states of the intermediate nucleus up to around 10 MeV. Suhonen and Civitarese \cite{Civitarese} have studied the two-neutrino double beta decay of $A=100$ nuclear systems within pnQRPA. In their calculation, they have considered the contribution of the intermediate $1^+$ states up to around 30 MeV. The advantage of pnQRPA is that it can take the Gamow Teller giant resonance (GTGR) region into account realistically.  In the previous shell model calculations for $^{48}$Ca \cite{Kostensalo, Horoi}, the cumulative NME  
	shows largest fluctuations  below 10 MeV energy corresponding to the intermediate $1^+$ states in $^{48}$Sc. However, in the case of the recent QRPA calculation, the second largest contribution is coming around 11 MeV, which is the giant resonance of $^{48}$Ca$\rightarrow^{48}$Sc \cite{Terasaki}. The relative phase of the GT contributions of the parent to intermediate nuclei and of the daughter to intermediate nuclei have not been measured \cite{Shimizu}, and hence its contribution to the $2\nu\beta\beta$ decay has not been quantified. Therefore, we feel that even with the limited model space considered in our large-scale shell model calculations, result should not be far from reality except for the contributions from the GTGR region.

	\begin{figure}[h]
		\begin{center}
			\includegraphics[width=110mm]{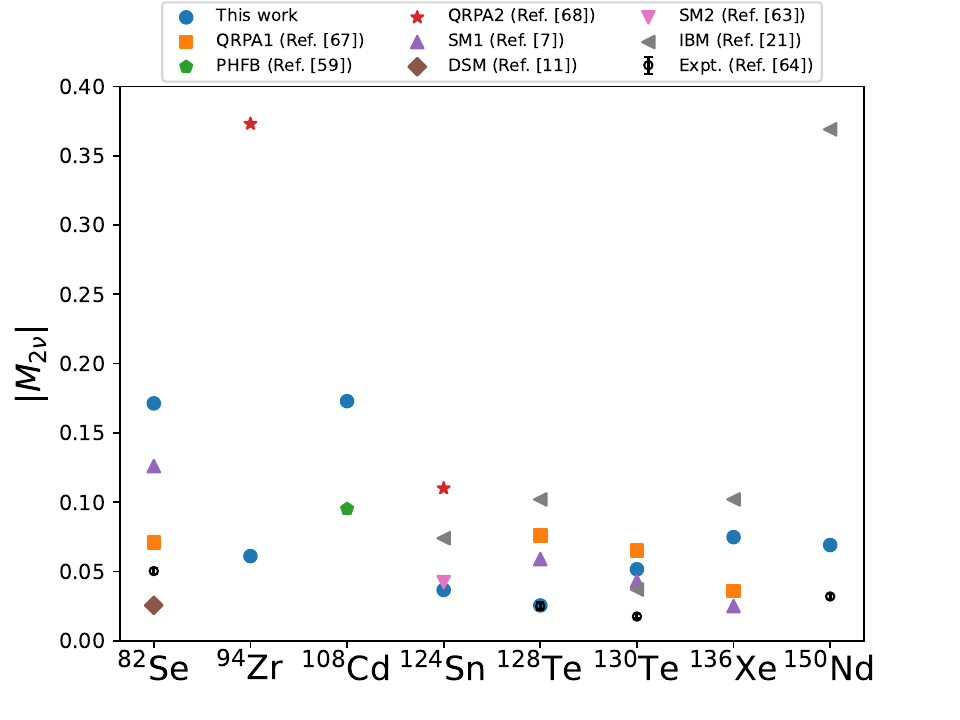}
		\end{center}
		\caption{\label{fig2} Comparison of shell model calculated NME with the different nuclear models available in the literature.}
	\end{figure}

	\begin{landscape}
		\begin{table*}
			\centering
			\caption{Comparison between shell model calculated and previously obtained NMEs ($|M_{2\nu}|$). The proton and neutron model space in the present work symbolize as $P$ and $N$, respectively. In the previous shell model calculations, '*' symbols denote that the $0g_{7/2}1d2s0h_{11/2}$ proton and neutron orbitals are included in the model space. The '$\dagger$' symbol corresponds that the $0f_{5/2}1p0g_{9/2}$ proton and neutron orbitals are involved.}
			\begin{tabular}{ccc|cccccccc}
				\hline 
				
				& \multicolumn{2}{c}{Present work}  &    \multicolumn{8}{c}{NMEs from Other works} \\

				\cline{2-3} 
				\cline{4-11}
				
				Isotope	& \makecell{Model space \\ (Interaction name)} & NME & \makecell{QRPA1 \\ \cite{Simkovic}} &  \makecell{PHFB \\ \cite{Raina}}  &  \makecell{QRPA2 \\ \cite{Suhonen2}}  &  \makecell{DSM \\ \cite{Sahu}} &  \makecell{IBM \\ \cite{Nomura}}  & \makecell{SM1 \\ \cite{Caurier1}} &  \makecell{SM2 \\ \cite{Horoi1}} &  \makecell{Expt. \\ \cite{Barabash1}} \\

				\hline
				
				{\bf $^{82}$Se}	& \makecell{$P\rightarrow 0f_{5/2}1p0g_{9/2}$ \\ $N\rightarrow 0f_{5/2}1p0g_{9/2}$ \\ (JUN45)} & 0.1713 & 0.071  & - & - & 0.0256 & -  & 0.126$^{\dagger}$ & - & 0.0503$^{+0.0020}_{-0.0018}$ \\
				
				\hline
				
				{\bf  $^{94}$Zr}	& \makecell{$P\rightarrow 0f_{5/2}1p0g_{9/2}$ \\ $N\rightarrow 0g_{7/2}1d2s$ \\ (GLEKPN)} & 0.0611 & - & - & 0.373 & - & - & - & - & - \\
				
				\hline
				
				{\bf  $^{108}$Cd}	& \makecell{$P\rightarrow 1p_{1/2}0g_{9/2}$ \\ $N\rightarrow 0g_{7/2}1d2s$ \\ (G-matrix)} & 0.1729 & - & 0.0952 & - & - & - & - & - & - \\
				
				\hline
				
				{\bf $^{124}$Sn}	& \makecell{$P\rightarrow 0g_{7/2}1d2s$ \\ $N\rightarrow 0g_{7/2}1d2s0h_{11/2}$ \\ (SN100PN)} & 0.0367 & - & - &  0.110 & - & 0.074 & - & 0.0423$^*$ & - \\
				
				\hline
				
				{\bf $^{128}$Te}	& \makecell{$P\rightarrow 0g_{7/2}1d2s$ \\ $N\rightarrow 1d2s0h_{11/2}$ \\ (SN100PN)} & 0.0255 & 0.076 & - & - & - & 0.102 &  0.059$^*$ & - & 0.0249$^{+0.0031}_{-0.0023}$  \\

				\hline

				{\bf $^{130}$Te}	& \makecell{$P\rightarrow 0g_{7/2}1d2s$ \\ $N\rightarrow 0g_{7/2}1d2s0h_{11/2}$ \\ (SN100PN)} & 0.0516 & 0.065  & - & - & - & 0.037 & 0.043$^*$ & - & 0.0175$^{+0.0016}_{-0.0014}$ \\

				\hline
				
				{\bf $^{136}$Xe}	& \makecell{$P\rightarrow 0g_{7/2}1d2s0h_{11/2}$ \\ $N\rightarrow 0g_{7/2}1d2s0h_{11/2}$ \\ (SN100PN)} & 0.0748 & 0.036  & - & - & - & 0.102 & 0.025$^*$ & - & - \\
				
				\hline
				
				{\bf $^{150}$Nd}	& \makecell{$P\rightarrow 0g_{7/2}2s0h_{11/2}$ \\ $N\rightarrow 0h_{9/2}0i_{13/2}$ \\ (KHHE)} & 0.0691 & - & - & - & - & 0.369 & - & - & 0.0320$^{+0.0018}_{-0.0017}$ \\

				\hline 
				
			\end{tabular}
			\label{NME}
		\end{table*}
	\end{landscape}

	\subsection{\bf Comparison of calculated NMEs}

	In Fig. \ref{fig2}, we have compared our shell model calculated NME with the previous results available in the literature using different nuclear models. The same data are also reported in Table \ref{NME}. For $^{82}$Se, the NME (0.071) from Ref. \cite{Simkovic} is less than half of our shell-model calculated value. However, the proton-neutron quasiparticle random-phase approximation (pnQRPA) calculated NME \cite{Suhonen2} for $^{94}$Zr has a larger value than our calculated result. Using this previously calculated NME 0.373 \cite{Suhonen2} for the minimum value of proton-neutron particle-particle interaction strength parameter ($g_{pp}$), we have calculated the half-life equals to $0.24\times10^{23}$ yr, which is smaller than the lower limit of recommended half-life (0.31$\sim$66)$\times 10^{23}$ yr \cite{Tretyak}. It shows that the shell model suggested NME for $^{94}$Zr is more compatible with predicting the accurate half-life than the previous calculation. In Ref. \cite{Raina}, the calculated NME for $2\nu$ double electron capture process in $^{108}$Cd using the projected Hartree-Fock-Bogoliubov (PHFB) model is approximately 45\% smaller than the shell-model calculated result. Our calculations predict the NME for $2\nu\beta\beta$ in $^{124}$Sn as 0.0369, which is close to the NME calculated in Ref. \cite{Horoi1}. The half-life prediction using our calculated NME may be useful for comparison with the upcoming experimental data. The shell model predicted NMEs for $2\nu\beta\beta$ decay in $^{128,130}$Te are in reasonable agreement with the NMEs calculated in the other works; our calculated NMEs are also quite similar to the previous works  \cite{Barabash1}, and \cite{Simkovic}, respectively. In case of $^{136}$Xe, shell model calculations suggest approximately two times larger NME (0.0748) than the QRPA calculated NME (0.036) \cite{Simkovic}, which predicts a half-life ($1.27\times10^{21}$), which is close to the average value $(2.18\pm0.05)\times10^{21}$ yr compared to the previous work. In the case of $^{150}$Nd,  we have calculated the NME considering just 75 intermediate $1^+$ states only because, computationally, it is challenging. However, our calculated NME (0.0691) is close to the earlier extracted NME (0.063$\pm$0.003) reported in Ref. \cite{Caurier1}. Rath $et$ $el.$ \cite{Rath} also calculated the NMEs for $^{150}$Nd as 0.033, 0.027, 0.032, and 0.027 using the PHFB wave functions generated with PQQ1, PQQHH1, PQQ2, and PQQHH2 parameterizations of effective two-body interaction, respectively. The extracted half-lives of $^{150}$Nd for $2\nu\beta\beta$ decay using these NMEs (as reported in Ref. \cite{Rath} ) are approximately four to six times larger than our shell model calculated value.
	
	\section{Summary and Conclusions} \label{section4}
	
	In the present work, large-scale shell-model calculations were carried out for the study of $2\nu\beta\beta$-decay of $^{82}$Se, $^{94}$Zr, $^{108}$Cd, $^{124}$Sn, $^{128}$Te, $^{130}$Te, $^{136}$Xe, and $^{150}$Nd.  We have calculated more intermediate $1^+$ states to get NME in comparison to most of the previously available results in the literature. For the calculation of NME of the $2\nu\beta\beta$ decay process, five types of shell-model effective interactions are used with different model spaces. Results of shell-model calculated NMEs and extracted half-lives for these nuclei are reported in Table \ref{half-life}. In case of $^{82}$Se, $^{94}$Zr, $^{128}$Te, $^{130}$Te, $^{136}$Xe, and $^{150}$Nd the calculated half-lives of $2\nu\beta\beta$-decaying nuclei are in good agreement with the experimental data. The half-life of $^{124}$Sn for $2\nu\beta\beta$ decay is not yet observed experimentally. Our prediction for $t_{1/2}^{2\nu}$ may be useful to compare with the upcoming experimental data. Apart from this, we have also analyzed the contribution of intermediate $1^+$ states in the variation of cumulative $2\nu\beta\beta$ NME.
	We aim to report results for $^{100}$Mo in the future, and at present, the shell model dimensions needed for this nucleus are too large for the computational facilities available to us. Also, with the good agreement obtained for the $2\nu$ DBD half-lives for the eight nuclei studied in this paper, we plan to carry out SM calculations for these nuclei for $0\nu$ DBD nuclear transition matrix elements in the near future.
	
	\section*{\uppercase{Acknowledgements}}
	
	We acknowledge financial support from SERB (India) under the research project CRG/2022/005167. 
	We would also like to thank the National Supercomputing Mission (NSM) for providing computing resources of ‘PARAM Ganga’ at the Indian Institute of Technology Roorkee, implemented by C-DAC and supported by the Ministry of Electronics and Information Technology (MeitY) and Department of Science and Technology (DST), Government of India.


	\bibliographystyle{apalike}

\end{document}